\documentclass{ws-procs975x65}            % default, citations in superscript
\usepackage{latexsym}
\usepackage{graphicx}
\usepackage{verbatim}
\usepackage{multirow}
\usepackage{amsmath}
\newcommand{\beq}{\begin{eqnarray}}
\newcommand{\eeq}{\end{eqnarray}}
\usepackage{mathrsfs}
\usepackage{float}
\begin{document}
\title{Bose Metal as a Disruption of the Berezinskii-Kosterlitz-Thouless Transition in 2D Superconductors}

\author{Philip W. Phillips$^*$ }

\address{Loomis Laboratory of Physics and Institute for Condensed Matter Theory, University of Illinois,\\
Urbana-Champaign, Il. 61801-3080\\
$^*$dimer@illinois.edu\\
}

\begin{abstract}
Destruction of superconductivity in thin films was thought to be a simple instance of Berezinskii-Kosterlitz-Thouless physics in which only two phases exist:  a superconductor with algebraic long range order in which the vortices condense  and an insulator where the vortex-antivortex pairs proliferate.  However, since 1989 this view has been challenged as now a preponderance of experiments indicate that an intervening bosonic metallic state obtains upon the destruction of superconductivity.    Two key features of the intervening metallic state are that the resistivity turns on continuously from the zero resistance state as a power law, namely $\rho_{\rm BM}\propto (g-g_c)^\alpha$ and the Hall conductance appears to vanish.  We review here a glassy model which is capable of capturing both of these features.  The finite resistance arises from three features.  First, the disordered insulator-superconductor transition in the absence of fermionic degrees of freedom (Cooper pairs only), is controlled by a diffusive fixed point\cite{CN} rather than the critical point of the clean system.  Hence, the relevant physics that generates the Bose metal should arise from a term in the action in which different replicas are mixed.  We show explicitly how such physics arises in the phase glass.  Second, in 2D (not in 3D) the phase stiffness of the glass phase vanishes explicitly as has been shown in extensive numerical simulations\cite{ky,kosterlitz1,kosterlitz2}.  Third, bosons moving in such a glassy environment fail to localize as a result of the false minima in the landscape.  We calculate the conductivity explicitly using Kubo response and show that it turns on as a power law and has a vanishing Hall response as a result of underlying particle-hole symmetry.  We show that when particle-hole symmetry is broken, the  Hall conductance turns on with the same power law as does the longitudinal conductance.  This prediction can be verified experimentally by applying a ground plane to the 2D samples.

\end{abstract}


\bodymatter

\section{Phenomenology}

Probably no other problem better exemplifies  the key topic of this conference, the Berezinskii-Kosterlitz-Thouless\cite{berezinskii,KT} (BKT) transition, than the insulator-superconductor transition\cite{hebard1,hebard2,paalanen628,G1989,mpaf} in thin films.  Key observations which helped place this transition within this framework are 1) a zero-resistive state below a critical value of the tuning parameter (either film thickness or magnetic field) interpreted as a condensation of vortex-antivortex excitations, 2) non-linear I-V vortex-antivortex excitations out of the zero-resistance state that give rise to the universal $V\approx I^{3}$ current-voltage\cite{Tsen2015} characteristics, and 3) exponential drop\cite{Tsen2015,liu2009} of the resistance below $H_{c2}$, indicative of thermal activation of vortex-antivortex motion with a binding energy of the BKT\cite{berezinskii,KT} form, $U(H)=U_0\ln H_0/H$ where $H_0\approx H_{c2}$ and $U_0$ the binding energy.  

However, a key feature which does not fit this scenario is the eventual leveling of the resistance for $T\ll U$.  Since this phase obtains below $H_{c2}$, the excitations are fundamentally bosonic, hence the Bose metal.  Within the standard XY\cite{mpaf} modeling of the BKT transition, this state of affairs  is not possible.  In this scenario, a metallic state only obtains at the critical point separating the ordered and disordered states and the quantum of resistance should be $h/4e^2$.   Indeed, the initial experiments\cite{hebard2,hebard1,paalanen628} seemed to be in agreement with the predictions\cite{mpaf} of the phase-only XY model that only at the critical point do bosons exhibit the  quantum of resistance of $h/4e^2$.  However, further experiments\cite{G1989,yazdani,valles628,vallesmetal} indicated that there is nothing special about the value of the resistance at the critical point, thereby calling into question the relevance or accuracy of the prediction of the phase-only model.   In fact since 1989\cite{ephron,yoon2006,park2017,iwasa,mason,masonthesis,vallesmetal,vallesgauge,liu2009,G2005,yazdani3,armitage2013,Tsen2015,thicknessbehnia}, a state with apparent finite $T\rightarrow 0$ resistivity appeared immediately upon the destruction of superconductivity.  Although the initial observations were derided as an artifact of failed refrigeration\cite{G1989}, the leveling of the resistance persisted in the magnetic-field tuned transition in MoGe\cite{ephron,mason,yazdani3}, Ta\cite{yoon2006,yoonTa}, InO$_x$\cite{armitage2013,kapsym}, and NbSe$_2$\cite{liu2009,Tsen2015}.  As mentioned above, the key contribution of the magnetic-field tuned data was to clarify that the intervening state occurred well below $H_{c2}$ in the regime where $T\ll U$, thereby requiring something beyond the classical physics of the BKT transition.  
\begin{figure}
\includegraphics[scale=0.65,angle=0]{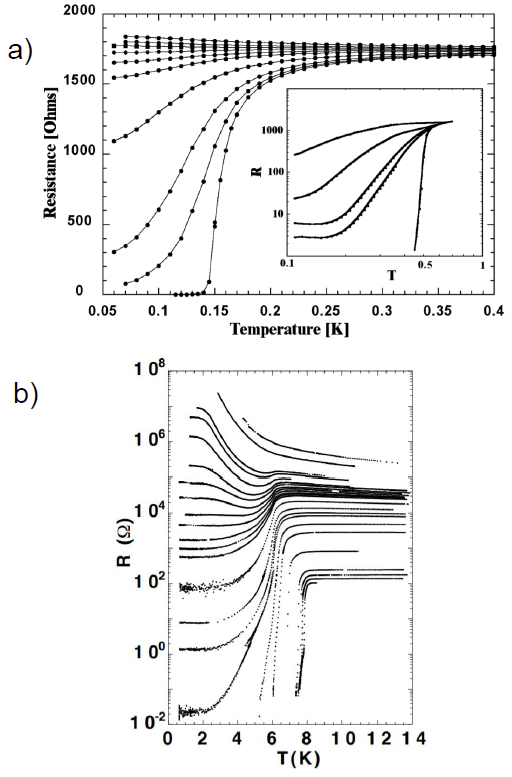}
\caption{Resistivity versus temperature for two different systems. a) Electrical resistance of MoGe thin film plotted vs temperature at B=0, 0.5, 1.0, 2.0,
3.0, 4.0, 4.4, 4.5, 5.5, 6 kG. The sample becomes a superconductor at 0.15 K in zero field but
for fields larger than about 4.4 kG the sample becomes insulating. At fields lower than this but
other than zero, the resistance saturates. The saturation behavior is better shown in the inset
for another sample with a higher transition temperature. The inset shows data for B= 0, 1.5, 2,
4,and 7kOe. At higher field, this sample is an insulator. Main figure reprinted from A. Yazdani
and A. Kapitulnik, Phys. Rev. Lett. 74, 3037 (1995), while the inset is from, Phys. Rev. Lett.
76. 1529 (1996).  b) Reprinted from C. Christiansen, L. M. Hernandez, and A. M. Goldman, Phys. Rev.
Lett., 88, 37004 (2002). Evolution of the temperature dependence of the resistance for a series
of Ga films. Film thicknesses range from 12.75 \AA to 16.67\AA and increases from top to bottom.
The leveling of the resistance once superconductivity is destroyed (zero resistance curves) is
not consistent with conventional wisdom. Note that the plateau value of the resistivity increases
as the distance from the superconducting phase increases. }
\end{figure}

More recent observations of the Bose metal in cleaner samples with either gate\cite{iwasa} or magnetic-field tuning\cite{Tsen2015} disclose three facts of the transition.  First, in the field-effect transistors\cite{iwasa} composed of ion-gated ZrNCl crystals, the superconducting state that ensues for gate voltages exceeding 4V is destroyed\cite{iwasa} for perpendicular magnetic fields as low as $0.05T$.  This behaviour was attributed\cite{iwasa} to weak pinning of vortices and hence the authors reach the conclusion that throughout most of the vortex state, be it a liquid or a glass, a metallic state obtains. This conclusion is particularly telling and of fundamental importance to the eventual construction of the theory of the metallic state.  Second, in  NbSe$_2$ an essentially crystalline material,  the resistance turns on\cite{Tsen2015} continuously as $\rho\approx (g-g_c)^\alpha$, where $g_c$ is the critical value of the tuning parameter for  the onset of the metallic state.  Similar results have also been observed in MoGe\cite{masonthesis}.   Third, in InO$_x$ and TaN$_x$, the Hall conductance is observed\cite{kapsym} to vanish throughout the Bose metallic state, thereby indicating that particle-hole symmetry is an intrinsic feature of this  state.  A similar vanishing of the Hall resistance below a critical value of the applied field was seen earlier by Paalanen, Hebard and Ruel on amorphous indium oxide films\cite{paalanen628}.  In strong support of this last claim are the recent experiments demonstrating that the cyclotron resonance vanishes in the Bose metallic state\cite{armitage}.  

While there have been numerous proposals for a Bose metal\cite{p1,p2,p3,p4,p5,dp2},  a state with a finite resistance at $T=0$ (demonstrated to exist in only one purely bosonic model\cite{dp2}), the new experiments highly constrain possible theoretical descriptions.  Of particular importance is the observation that in the clean samples\cite{iwasa}, the vortex state that ensues once the superconducting state is destroyed is metallic!   This appears to be in potential conflict with the vortex glass state being a superconductor\cite{fisher,vortexglass}.    In fact, the broad observation of a metallic state in 2D samples, be they disordered or quite clean, points to a re-examination of the ultimate fate of vortex states in 2D.  It is precisely this that we do here as the models my group proposed several years ago\cite{dp1,dp2,jw1} are all based on glassy vortex states in which the conductivity was shown to be finite from an explicit calculation of the conductivity from the Kubo formula.  In the collision-dominated (or hydrodynamic) regime, the resistivity has a finite value and turns on as $\rho\approx (g-g_c)^\alpha$, as highlighted in the experiments on NbSe$_2$\cite{Tsen2015}.  While questions\cite{stiffness} regarding the phase stiffness of the phase glass have been raised, numerical simulations all indicate\cite{ky,kosterlitz1,kosterlitz2} that the energy to create a defect in a 2D phase or gauge glass scales as $L^{\theta}$, where $\theta=-0.39$.  Hence, the stiffness is non-existent.  In 3D\cite{ky,kosterlitz1,kosterlitz2}, $\theta>0$ and a stiffness obtains.  As such glass states are candidates to explain the vortex glass\cite{fisher,vortexglass},  that $\theta<0$ is consistent with the experimental finding\cite{iwasa} in ion-gated ZrNCl, an extreme 2D system, that the resultant vortex state is indeed metallic and not a true superconductor.  

\section{Preliminaries}

In the 80's and '90's, the leading candidate to explain failed superconductivity was the resistively shunted Josephson junction array model\cite{chakrav,chakravdiss,otterlo}.  All such models are based on a propagator of the form,
\beq
G_0=(k^2+\eta|\omega_n|+m^2)^{-1},
\eeq
in which the Ohmic dissipative term, $|\omega|$ is an attempt to model the normal electrons.  Despite the Ohmic term, all such models yield\cite{otterlo,dp2000} either insulating or superconducting states and hence are not applicable to the metallic state.   In computing the conductivity of these models, it is important to note\cite{DS1997} that the conductivity in the vicinity of the transition region is a universal function of the form, $\sigma_Q f(\omega/T)$ where $f$ is a monotonically decreasing function of the frequency, $\omega$, and the temperature, $T$.  The experiments correspond to the limiting procedure $\lim_{T\rightarrow 0}\lim_{\omega\rightarrow 0}$, that is to $f(0)$ not the inverse limit where $f(\infty)$ enters. The physics of $f(0)$ is pure hydrodynamics in which it is collisions of the quasiparticle excitations of the order parameter that regularize the conductivity.  Explicit computation of the conductivity in the resistivively shunted Josephson junction array model yielded\cite{dp2000}  that although dissipation can drive an intermediate region in temperature where the resistivity levels, ultimately at $T=0$ a superconductor obtains.  Hence, dissipation alone cannot drive the Bose metal.  

Before we present the glassy model that has the extra ingredient, it is instructive to look at a simpler model which illustrates the power of the hydrodynamic\cite{dalidovich} approach.  Our system consists of an array of Josephson junctions,
we coarse-grain over the phase associated with each junction and
hence use as our starting point the Landau-Ginzburg action,
\beq\label{gaussian}
F[\psi]&=&\sum_{\vec k,\omega_n}(k^2+\omega_n^2+m^2)
|\psi(\vec k,\omega_n)|^2\nonumber\\
&&+\frac{U}{2N\beta}\sum_{\omega_1,...,\omega_4; \vec k_1,...,\vec k_4}
\delta_{\omega_1+\cdots\omega_4,0} \delta_{\vec k_1+ \cdots \vec k_4,0}\nonumber\\
&&\psi_\nu(\omega_1,\vec k_1)\psi_\nu(\omega_2,\vec k_2)
\psi_\mu(\omega_3,\vec k_3)\psi_\mu(\omega_4,\vec k_4),
\eeq
where $\psi(\vec r,\tau)$ is the complex Bose order parameter
whose expectation value
is proportional to $\langle\exp(i\phi)\rangle$, where $\phi$ is the phase
of a particular junction.  The summation in the action
over discrete Matsubara frequencies, $\omega_i=2\pi n_i T$,
and integration over continuous wavevectors, $\vec k$, is assumed. 
The parameter $m^2$ is 
the inverse square of the correlation length. In writing the action in
this fashion, we have already included the one-loop renormalization
arising from the quartic term.  In the quantum-disordered regime,
$m\gg T$ and hence it is the quantum fluctuations that dominate
the divergence of the correlation length.   

This model has two phases, a superconductor for $m^2<0$ and an insulator for $m^2>0$.   The conductivity in the insulator is expected to vanish.  But in fact it does not precisely because the resistivity in the quantum disordered regime is mediated by finite temperature collisions between the bosons.  Such events are exponentially activated.  However, their lifetime is also exponentially long as can be seen from an explicit calculation\cite{dalidovich}.  To recount the calculation, we note that the central quantity appearing
in the collision integral is the polarization function
\beq
\label{polariz}
\Pi(\vec q ,i\Omega_m)=T\sum_n \int \frac{d^2p}{(2\pi)^2} 
G_0(\vec p +\vec q, \omega_n +\Omega_m) G_0(\vec p, \omega_n)\nonumber\\
\eeq 
where the bare field propagator $G_0(\vec p, \omega_n)=
(p^2+\omega_n^2+m^2)^{-1}$. As it is the imaginary
part of $\Pi$ that is required in the collision integral, we must perform an 
analytical continuation
$\Omega_n\rightarrow -i\Omega_n -\delta$ with $\delta$ a positive
infinitesimal.   It is the polarization function that appears explicitly in the scattering time,
\beq
\label{gentau}
\frac{1}{\tau_{\vec k}} &=& \frac{1}{2(2\pi )^2} \left[   
  \int \frac{d^2 q}{\epsilon_{\vec{q}+\vec{k}}\epsilon_{\vec k}}
  \left( {\rm Im}\frac{1}{\Pi (\vec{q}, \epsilon_{\vec{q}+\vec{k}}-
   \epsilon_{\vec k})} \right) n(\epsilon_{\vec{q}+\vec{k}}-
   \epsilon_{\vec k}) \right.\nonumber\\
  && \left. + \int \frac{d^2 q}{\epsilon_{\vec q}\epsilon_{\vec k}}
    \left( {\rm Im}\frac{1}{\Pi (\vec{q}+\vec{k}, \epsilon_{\vec q}
    +\epsilon_{\vec k})}
    \right) n(\epsilon_{\vec{q}}) \right],
\eeq
which is related to the conductivity through
\beq
\label{sigma}
\sigma=2\frac{(e^\ast)^2}{\hbar}\int\frac{d^2k}{(2\pi)^2}\frac{k_x^2}
{\epsilon_{\vec k}^2}
\tau_{\vec k}\left(-\frac{\partial n(\epsilon_{\vec k})}
{\partial\epsilon_{\vec k}}\right).
\eeq
The essence of our central result is
that to leading order in $T/m$, the inverse relaxation time
$1/\tau_{\vec k}$ is momentum-independent and given by
\beq
\label{relaxtime}
 \frac{1}{\tau}=\pi Te^{-m/T}.
\eeq
Substitution of this expression into Eq. (\ref{sigma}) leads to the mutual 
cancellation of the exponential factors yielding the remarkable result
\beq
\label{mainresult}
 \sigma(T\rightarrow 0)=\frac{2}{\pi} \frac{4e^2}{h}.
\eeq
It is curious to note\cite{fradkin,plee} that a similar cancellation of exponential
 factors (from the mean free path and the density of states)
arises in the context of the quasiparticle conductivity in a dirty d-wave superconductivity
yielding the identical numerical prefactor $2/\pi$.  In essence, the insulator is a metal because the mean-free path of the bosonic excitations is exponentially small with the same factor that enters the population of bosonic excitations.  Since it is the product of the scattering time and the population that leads to the conductivity, the result must be finite. While this result is interesting, this metal state is quite fragile as it is destroyed by any perturbation.  Hence, the answer to the experiments lies elsewhere.

\section{Bose Metal}
As shown previously\cite{spivakkiv}, any amount of dirt in a 2D superconductor induces $\pm J$ disorder, $J$ the Josephson coupling.  Consequently, a disordered superconductor is closer to a disordered XY model  rather than a dirty superfluid.  Justifiably, the  starting point for analyzing the experiments is the disordered XY model,
\beq\label{HJ}
H=-E_C\sum_i\left(\frac{\partial}{\partial\theta_i}\right)^2-
\sum_{\langle i,j\rangle} J_{ij}\cos(\theta_i-\theta_j), 
\eeq
with random Josephson couplings $J_{ij}$ but fixed
on-site energies, $E_C$.   The phase of each island
is $\theta_i$. If the Josephson couplings are chosen from a distribution
with zero mean, only two phases are possible: 1) a glass arising from the distribution of positive and negative $J_{ij}'s$ and 2) a disordered paramagnetic state.  A superconducting phase obtains if the distribution
\beq\label{distj}
P(J_{ij})=\frac{1}{\sqrt{2\pi
J^2}}\exp{\left[-\frac{(J_{ij}-J_0)^2}{2J^2}\right]}
\eeq
of $J_{ij}'s$ has non-zero mean, $J_0$, and $J$ the variance.
 To distinguish between the phases, it is expedient
to introduce\cite{dpprb} the set of variables ${\bf S}_i=(\cos\theta_i,
\sin\theta_i)$
which allows us to recast the interaction term
in the random Josephson Hamiltonian as a spin problem
with random magnetic interactions,
 $\sum_{\langle i,j\rangle}J_{ij}{\bf S}_i \cdot {\bf S}_j$.
Let $\langle ...\rangle$ and $[...]$ represent
averages over the thermal degrees of freedom
and over the disorder, respectively.  Integrating over the random interactions
will introduce two auxilliary fields
\beq
 Q_{\mu\nu}^{ab}(\vec k,\vec k',\tau,\tau')=\langle
S_\mu^a(\vec k,\tau)S_\nu^b(\vec k',\tau')\rangle
\eeq
and $\Psi^a_\mu(\vec k,\tau)=\langle S^a_\mu(\vec k,\tau)\rangle$,
respectively. The
superscripts represent the replica indices. A non-zero value of
$\Psi^a_\mu(\vec k,\tau)$ implies phase ordering of
the charge $2e$ degrees of freedom.
For quantum spin
glasses, it is the diagonal elements of the Q-matrix
$D(\tau-\tau')=\lim_{n\rightarrow 0}\frac{1}{Mn}\langle
Q^{aa}_{\mu\mu}(\vec k,\vec k',\tau,\tau')\rangle$ in the limit that
$|\tau-\tau'|\rightarrow\infty$ that serve as the effective
Edwards-Anderson spin-glass order parameter\cite{sachdev,huse,bm}
within Landau theory. The free energy per replica
\beq\label{fen}
&&{\cal F}[\Psi,Q]={\cal F}_{\rm SG}(Q)+
\sum_{a,\mu, k,\omega_n}(k^2+\omega_n^2+m^2)
|\Psi_\mu^a(\vec k,\omega_n)|^2 \nonumber\\
&&-\frac{1}{\kappa t}\int d^d x\int d\tau_1
d\tau_2\sum_{a,b,\mu, \nu}\Psi_\mu^a(x,\tau_1)
\Psi^b_\nu(x,\tau_2)
Q_{\mu\nu}^{ab}(x,\tau_1,\tau_2)\nonumber\\
&&+U\int d\tau\sum_{a,\mu}\left[\Psi_\mu^a(x,\tau)\Psi_\mu^a(x,\tau)\right]^2,
\eeq
consists of a spin-glass part which is a third-order functional
of the $Q-$ matrices discussed  
previously\cite{dp2,sachdev}, the $\Psi_\mu^a$ terms that describe 
the charge 2e condensate and the term
which couples the charge and glassy degrees of freedom. The parameters,
$\kappa$, $t$ and $U$ are the standard coupling constants in
a Landau theory and $m^2$ is the inverse correlation length.  The essential aspect of the quantum rotor spin glass is that the saddle point solution
for the corresponding action is minimized by a solution of the form
\beq\label{qm}
Q_{\mu\nu}^{ab}(\vec k,\omega_1,\omega_2)&=&\beta(2\pi)^d\delta^d(k)
\delta_{\mu\nu}\left[
D(\omega_1)
\delta_{\omega_1+\omega_2,0}\delta_{ab}
+\beta\delta_{\omega_1,0}\delta_{\omega_2,0}q^{ab}
\right],
 \eeq
where the diagonal elements are given by
\beq
D(\omega)=-\sqrt{\omega^2+\Delta^2}/\kappa,
\eeq
with $\kappa$ a coupling constant in the Landau free energy for the spin glass.
The diagonal elements of the $Q$-matrices describe the excitation spectrum.
Throughout the glassy phase, $\Delta=0$ and hence the spectrum is ungapped and given by
$D(\omega)=-|\omega|/\kappa$. The linear dependence on $|\omega|$ arises
because the correlation function $Q^{aa}_{\mu\mu}(\tau)$ decays\cite{sachdev,huse,bm} as
 $\tau^{-2}$.
 This dependence results in a fundamental change in the dynamical critical exponent from
$z=1$ to $z=2$. 

To compute the conductivity in the glassy phase, we note that near  the spin-glass/superconductor boundary, $m^2$ should be regarded as 
the smallest parameter. Hence, it is the fluctuations of $\Psi_\mu^a$ rather 
than those of $Q^{ab}$ that dominate.  In this regard, we recall the work of 
Chamon and Nayak\cite{CN} who noted that that the disorder at the insulator-superconductor phase transition
drives the critical behaviour away from that of the clean system towards a diffusive fixed point in which the critical resistance
vanishes.  They then advocate that the resistivity should turn on continuously from zero as a power law.  It is precisely this behaviour that 
we prove here.  

To compute the conductivity, we focus on the part of the free energy,
\beq\label{fgauss}
{\cal F_{\rm gauss}}&=&\sum_{a,\vec k,\omega_n}
(k^2+\omega_n^2+\eta |\omega_n|+m^2)
|\psi^a(\vec k,\omega_n)|^2 \nonumber\\
&&-\beta q\sum_{a,b,\vec k,\omega_n} \delta_{\omega_n,0}
\psi^a(\vec k,\omega_n)[\psi^b(\vec k,\omega_n)]^{\ast},
\eeq
governed by the fluctuations of the superconducting order parameter at the Gaussian level.
In the above action, we introduced the effective dissipation 
$\eta=1/(\kappa^2 t)$ and rescaled $q\rightarrow q\kappa t$.  As is evident in this action, disorder appears explicitly as
a mixing between the replicas.  It is from this term that fundamentally new physics arises and the origin of the transition to the diffusive fixed point.
The new physics is captured by the  associated Gaussian propagator 
\beq\label{prop}
G^{(0)}_{ab}(\vec k,\omega_n)=G_0 (\vec k,\omega_n)\delta_{ab}+
\beta G_0^2(\vec k,\omega_n)q\delta_{\omega_n,0}
\eeq
which is obtained by inverting the free energy in ultrametric space\cite{zinjust}  in the $n\rightarrow 0$ 
limit\cite{zinjust} with
$G_0 (\vec k,\omega_n)=(k^2+\omega_n^2+\eta|\omega|+m^2)^{-1}$. 
The first term in
Eq. (\ref{prop}) is the standard Gaussian propagator in
the presence of Ohmic dissipation. The Ohmic dissipative term
in the free-energy arises from the diagonal elements of the 
$Q-$ matrices.  However, it is the $q-$dependent term
in the Gaussian free energy, the last term in Eq. (\ref{fgauss}), that is new
and changes fundamentally
the form of the propagator.  Because of the $\delta_{\omega_n,0}$ factor
in the second term in the free energy, the propagator now contains 
a frequency-independent part, $\beta G_0^2(\vec k,\omega_n=0) q$. 

To compute the conductivity, we use the generalization\cite{herbut} of the Kubo formula for
a replicated action and obtain for the conductivity
\beq
\sigma(i\omega_n)&=&\frac{2(e^*)^2}{n\hbar\omega_n}T\sum_{a,b,\omega_m}
\int \frac{d^2k}{(2\pi)^2} 
\left[G^{(0)}_{ab}(\vec k,\omega_m)\delta_{ab}\right.\nonumber\\
&&\left. -2 k_x^2 G_{ab}^{(0)}(\vec k, \omega_m)G_{ab}^{(0)}
(\vec k,\omega_m+\omega_n)\right].
\eeq
The conductivity contains
two types of terms.  All terms not proportional to $q$ have been 
evaluated previously\cite{dp2000,dalidovich} and vanish as $T\rightarrow 0$. The terms
proportional to $q^2$ vanish in the limit $n\rightarrow 0$.  The only
terms remaining are proportional to $q$ and yield after an appropriate 
integration by parts
\beq
\sigma(i\omega_n)=\frac{8qe^{\ast 2}}{\hbar\omega_n}\int
\frac{d^2k}{(2\pi)^2}k_x^2G_0^2(\vec k,0)\left[ G_0(\vec k,0)-
G_0(\vec k,\omega_n)\right].\nonumber
\eeq
The momentum integrations are straightforward and yield
\beq
\sigma(\omega=0,T\rightarrow 0)=\frac{8e^2}{\hbar}\frac{q\eta}{2 m^4}
\eeq
a temperature-independent value for the conductivity as $T\rightarrow 0$.  
The dependence on
$q$ and $\eta$ imply that dissipation alone is insufficient
to generate a metallic state.  What seems to be the case
is that a bosonic excitation moving in a dissipative
environment in which many false minima exist does not localize because
it takes an exponentially long amount of time to find the ground state.
This is the physical mechanism that defeats localization in
a glassy phase.   From the dependence on $m^4$, we see clearly that the resistivity turns on as a power law
\beq
\rho\approx (g-g_c)^{2z\nu}
\eeq
as is seen experimentally\cite{Tsen2015,masonthesis} and consistent with the Chamon/Nayak\cite{CN} work that at the superconductor-insulator transition, the resistivity should turn on continuously from zero not $h/4e^2$.   As shown elsewhere\cite{dp2}, the finite resistivity obtained here is robust to the quartic interactions in Eq. (\ref{fen}).  Hence, the metallic state is not an artifact of the Gaussian approximation.  Quantifying how this exponent changes as a function of disorder is of utmost importance as it would determine if all observations of the Bose metal lie in the same universality class.  

Aside from the turn-on of the resistivity, the phase glass can also explain the apparent vanishing Hall response in the metallic state.  We have recently computed the Hall conductance\cite{Julian2018} and found it to vanish as a result of the inherent particle-hole symmetry in this model.  Using a model in which the glassy degrees of freedom were generated from a random magnetic field in the form, $\cos(\theta_i - \theta_j - A_{ij})$ (where $A_{i j} =  e^*/\hbar\int_i^j \textbf{A} d\textbf{l}$, ($e^* = 2e$))  rather than the constant term in Eq. (\ref{HJ}) Away from the particle-hole symmetric point, the Hall conductance turns on as
\beq
\sigma_H({i\omega_\nu}) = \frac{\lambda q (e^{*}m_H^2)^2}{ \hbar m^4 } (\frac{2}{x} - \frac{\Psi(1,\frac{x+2}{2x})}{x^3} )
\eeq
where $x=\frac{m^2_H}{m^2}$, $m_H^2 = \frac{e^*}{c \hbar}B$, and $\Psi(1,x)$ is the first digamma function.  The compensating effect on the diagonal conductivity is
\beq
\sigma_{xx}({i\omega_\nu}) = \frac{\eta q (e^{*}m_H)^2}{ \hbar m^4 } (\frac{2}{x} - \frac{\Psi(1,\frac{x+2}{2x})}{x^3} ).
\eeq
Hence, both have identical trends in a magnetic field.  This prediction should serve as a guide to further experiments.   

\section{Final Remarks}

In conclusion, all current experiments on the destruction of superconductivity in thin films in 2D can be understood within a glassy model as the intermediary before the onset of the insulating state.  An analogy which might be helpful here is with the Bose-Hubbard model.  In  the absence of disorder, this model admits a direct transition from a superfluid to a Mott insulator.  However, in the presence of disorder, a Bose glass\cite{fisherbh} phase intervenes disrupting the direct transition to the Mott insulator.  The work presented here implies a similar trend is manifest in the charged case as well.   The falsifiable prediction for the turn-on of the Hall conductance can be confirmed by ground-plane experiments and should offer a new window into the true nature of the ground state of the Bose metal. 

{\bf Acknowledgements:}  This paper is largely a review of the previous works on this topic I co-authored with D. Dalidovich and J. May-Mann.  I thank Steve Kivelson for pointing out the references on the numerical simulations on the stiffness in the 2D spin glass.  This works was funded by NSF-DMR-1461952.

%\bibliography{bmetalfinal}
%\bibliographystyle{ws-procs975x65}
\end{document}